# Generalized Conditional Displacement

**Shiran Even-Haim[1], Asaf A. Diringer[2], Ron Ruimy[1], Gefen Baranes[3], Alexey Gorlach[4], Shay Hacohen-Gourgy[2], Ido Kaminer[1*]**

[1]*Faculty of Electrical & Computer Engineering, Technion - Israel Institute of Technology, Israel*

[2]*Faculty of Physics, Technion-Israel Institute of Technology, Israel*

[3]*Department of Physics, Massachusetts Institute of Technology, Cambridge, Massachusetts 02139, USA*

[4]*Solid State Institute, Technion-Israel Institute of Technology, Haifa 32000, Israel*

[*]*kaminer@technion.ac.il*

**Conditional displacement with a qubit ancilla is a critical component in continuous-variable error correction protocols. We present the generalized conditional displacement operator, conditioned on a *qudit* ancilla, and explore potential implementations. We show how this operator can be used to enhance error correction with Gottesman-Kitaev-Preskill (GKP) codes.**

## Introduction

Quantum error correction (QEC) is essential for reaching large-scale quantum information processing schemes. The growing interest in quantum computing motivates further developments of building blocks for QEC protocols. A promising approach in this direction is encoding quantum information in continuous variables (CV), i.e., in the states of *harmonic oscillators*. This approach relies on bosonic codes such as binomial codes [1], cat codes [2], and the Gottesman-Kitaev-Preskill (GKP) codes [3]. The complexity of controlling such a CV system usually benefits from combining it with another discrete-variables (DV) system to form a hybrid CV-DV system [4], offering the strengths of both CV and DV systems, as recently demonstrated in groundbreaking experiments with trapped ions and superconducting circuit QED [5–7].

All the demonstrations and most of the theory works in this area rely on the same fundamental building block: a quantum gate that performs displacement on the harmonic oscillator, conditioned on the state of a two-level ancilla, a.k.a conditional displacement (CD). In continuous variables quantum information processing, the CD operation is a “Swiss army knife”: it controls, measures, and stabilizes GKP states [8].

The CD operation combined with ancilla qubit rotations are the two critical building blocks for QEC and universality [8], as the CD operations can compose the code's stabilizers. Furthermore, CD enables the implementation of gate teleportation [9], which is crucial in quantum computation and communication applications.

Here, we analyze a generalization of the CD operation, called *generalized CD*, operating on an ancilla *qudit* rather than a qubit. The idea is to exploit the larger Hilbert space of the multi-level qudit and use the extra degrees of freedom to gain information faster with fewer operations. We show that using the generalized CD to stabilize and create GKP states is natural since it operates simultaneously on both symmetry axes of the GKP's grid structure. Employing generalized CD in QEC schemes for GKP enables more efficient stabilization and creation of GKP states.

We also discuss different interaction-Hamiltonians with different ancilla systems that generate the generalized CD gate. Each interaction-Hamiltonian either directly uses a qudit or approximates the generalized CD with a specific qudit encoding. We then propose and discuss different experimental implementations.

## Results

### Ancilla qudit for conditional displacement (CD) on harmonic-oscillator states

Displacement of a harmonic oscillator conditioned on a two-level system, a qubit, is defined by:

$$\mathrm{CD}_2(\alpha) \equiv D(\alpha Z) = |0\rangle\langle 0| \otimes D(\alpha) + |1\rangle\langle 1| \otimes D(-\alpha), \tag{1}$$

Where $Z$ is the qubit Pauli matrix and $D(\alpha) = e^{\alpha\hat{a}^\dagger - \alpha^*\hat{a}} = e^{-iRe(\alpha)\sqrt{2}\hat{p} + iIm(\alpha)\sqrt{2}\hat{q}}$ is the phase-space displacement, where $\hat{a}$ and $\hat{a}^\dagger$ are the annihilation and creation operators for the harmonic oscillator, and $\hat{q}$ and $\hat{p}$ are the conjugated position and momentum operators, such that $[q, p] = i$.

We propose to condition the displacement on a $d$-level ancilla qudit. For the case of a qudit of $d$-levels, the Pauli operators can be generalized to the non-Hermitian Heisenberg-Weyl qudit operators [10]: $\bar{Z}_d = \sum_{s=0}^{d-1} \omega_d{}^s |s\rangle\langle s|$ , $\bar{X}_d = \sum_{s=0}^{d-1} |s \oplus 1\rangle\langle s|$, where $d$ is the qudit dimension and $\omega_d = e^{i2\pi/d}$. The discrete qudit quadratures states $|s\rangle$ and $|m\rangle$ are the $\bar{Z}_d$ and $\bar{X}_d$ eigenstates [11]: $\bar{Z}_d|s\rangle = \omega_d{}^s|s\rangle$, $\bar{X}_d|m\rangle = \omega_d{}^{-m}|m\rangle$. The discrete quadrature operator is $\hat{s} = \sum_{s=0}^{d-1} s|s\rangle\langle s|$, and the dual quadrature operator

is $\hat{m} = \sum_{m=0}^{d-1} m|m\rangle\langle m|$ , with the Fourier-transform relations of: $|m\rangle = \frac{1}{\sqrt{d}}\sum_{s=0}^{d-1} {\omega_d}^{sm}|s\rangle = \frac{1}{\sqrt{d}}\sum_{s=0}^{d-1} {\bar{Z}_d}^{\ m}|s\rangle$, $|s\rangle = \frac{1}{\sqrt{d}}\sum_{m=0}^{d-1} {\omega_d}^{-sm}|m\rangle = \frac{1}{\sqrt{d}}\sum_{m=0}^{d-1} \bar{X}_d^{s}|m\rangle$ . The qudit operators can be written as translations in the discrete quadratures: $\bar{Z}_d = \omega_d^{\hat{s}}$, $\bar{X}_d = \omega_d^{-\hat{m}}$. Using these operators, we introduce the generalized CD operator for the joint Hilbert space of the qudit and the harmonic oscillator (Figure 1b):

$$\mathrm{CD}_d(\alpha) \equiv D(\alpha \bar{Z}_d) = e^{\alpha \bar{Z}_d \hat{a}^\dagger - \alpha^* \bar{Z}_d^\dagger \hat{a}} = \sum_{s=0}^{d-1} |s\rangle\langle s| \otimes D(\alpha \omega_d^s). \tag{2}$$

We note that there are other ways to generalize the CD operator. Instead of extending from qubit to qudit, it is also possible to extend to higher spins, based on the interaction of symmetric states (high-spin states) with the harmonic oscillator [12].

**Creating rotational codes**

The $d$-legged cat state with the generalized parity $m$ mod $d$ is defined by [13]:

$$|C_d^m(\alpha)\rangle = \frac{1}{\sqrt{dN_m(\alpha)}} \sum_{s=0}^{d-1} \omega_d^{-sm} |\alpha \omega_d^s\rangle, \tag{3}$$

where $N_m(\alpha) = \sum_{s=0}^{d-1} \omega_d^{-sm} e^{(\omega_d^s - 1)\alpha^2}$ is the normalization factor. One can choose an encoding of logical states as:

$$|0\rangle_L = \left|C_d^{m=0}(\alpha)\right\rangle, \qquad |1\rangle_L = \left|C_d^{m=\lfloor d/2 \rfloor}(\alpha)\right\rangle. \tag{4}$$

This code is quantum error correctable [14] for Fock space shifts and phase space rotations. For Fock space shifts, the code distance is $\lfloor d/2 \rfloor$. Therefore, correctable shifts are up to $\left\lfloor \frac{\lfloor d/2 \rfloor - 1}{2} \right\rfloor$. Similarly, the code has $\frac{4\pi}{d}$ rotation symmetry for phase-space rotations, and the code distance is $\frac{2\pi}{d}$. Therefore, correctable rotations are up to $\frac{\pi}{d}$.

Applying the generalized CD on the initial state of the ancilla qudit in the $\bar{X}_d$ eigenstate $|m = 0\rangle$ and measuring the $\hat{m}$ quadrature of the ancilla qudit will apply the following Kraus operator on the harmonic oscillator:

$$\langle m|\mathrm{CD}_d(\alpha)|m = 0\rangle = \frac{1}{d} \sum_{s=0}^{d-1} {\omega_d}^{-sm} D(\alpha \omega_d^s). \tag{5}$$

Therefore, an initial state of the oscillator in the vacuum $|0\rangle$ state yields:

$$\mathrm{CD}_d(\alpha)(|m=0\rangle \otimes |0\rangle) = \frac{1}{\sqrt{d}} \sum_{m=0}^{d-1} \frac{1}{\sqrt{N_m(\alpha)}} |m\rangle \otimes |C_d^m(\alpha)\rangle. \quad (6)$$

Measuring the $\hat{m}$ quadrature of the ancilla qudit will leave the harmonic oscillator in the $d$-legged cat state $|C_d^m(\alpha)\rangle$ (Figure 1c), as first proposed in a free electron implementation of the CD operation [15, 16]. The probability of measuring each $|m\rangle$ state in the ancilla is $\frac{1}{dN_m(\alpha)}$.

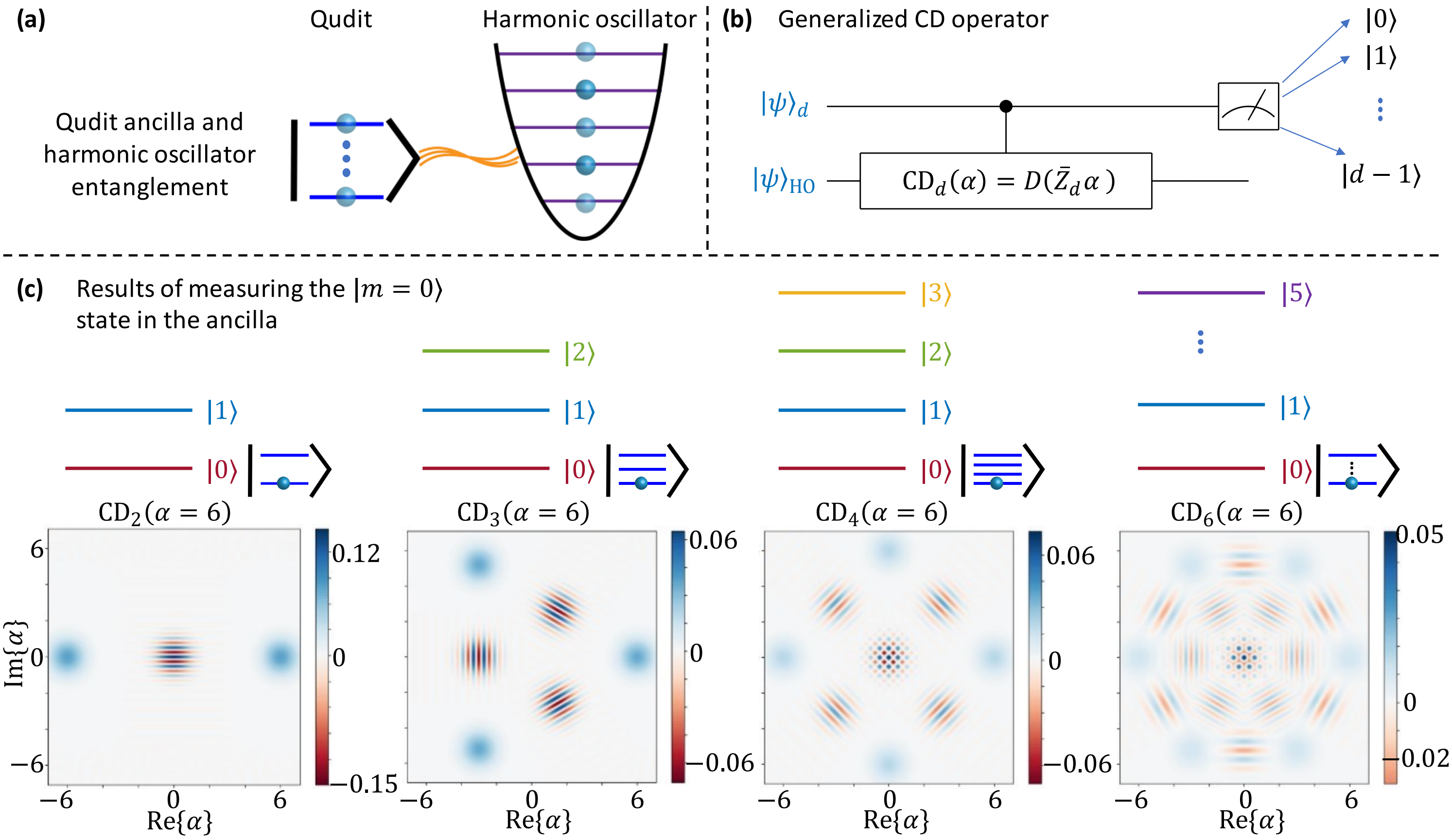

**Figure 1. Introducing the generalized conditional displacement (CD) operator: (a-b)** Entangling a $d$-level qudit and a harmonic oscillator, implementing Eq. (2). **(c)** Wigner representations after a generalized CD and qudit measurement in the $\bar{X}_d$ basis.

### Quantum error correction (QEC) with the generalized CD

The generalized CD operator can improve the stabilization and QEC protocols for the grid-state (GKP) code. This section will focus on the canonical square grid-state code defined by the abelian stabilizer group generated by $\hat{S}_X = D(l)$, $\hat{S}_Z = D(il)$, where $l = \sqrt{2\pi}$.

Ideal GKP code

The standard approach for GKP stabilization is measuring the stabilizer generators, i.e., measuring the modular quadratures a.k.a the Zak basis [17], $\hat{q} \bmod \sqrt{\pi}$ and $\hat{p} \bmod \sqrt{\pi}$ [18]. These measurements reveal the errors that can be corrected with a displacement. Idealized measurements of the modulo quadratures can be approximated to physical measurements with finite precision. Ref. [19] showed an approximation for these measurements using phase estimation with an ancilla qubit. Each round of phase estimation starts with a preparation of the ancilla in $|+\rangle$, followed by the CD operator with the displacement equal to the measured quadrature's stabilizer generator. The ancilla is then measured, revealing one bit of information used to displace the oscillator with a positive or negative fixed value, according to the measurement result. The overall stabilization process is therefore comprised of alternating rounds, split into stabilization in the $q$ quadrature and in the $p$ quadrature.

In contrast, using qudits with the generalized CD operator reveals more bits of information per phase estimation round. Since the stabilizer generators commute, measuring both modular quadratures simultaneously is equivalent to measuring each quadrature separately. Therefore, using the generalized CD operator with $d = 4$ can project onto an eigenstate of both generators simultaneously, stabilizing the oscillator state into the GKP manifold in half the required steps and giving faster convergence to the manifold.

In order to show the stabilization process using the generalized CD operator, we recall the conventional phase estimation with an ancilla qubit from [19]. Each round of phase estimation starts with the ancilla in $|+\rangle$, followed by the $\mathrm{CD}_2$ operator, and finally, the ancilla is measured along $e^{\mp\frac{i\pi}{4}Z}|+\rangle$. We evaluate the two Kraus operators acting on the oscillator associated with the measured ancilla qubit:

$$M_{\pm}(\alpha) = \langle +|e^{\pm\frac{i\pi}{4}Z}\mathrm{CD}_2(\alpha)|+\rangle = \frac{1}{2}\left(e^{\pm i\frac{\pi}{4}}D(\alpha) + e^{\mp i\frac{\pi}{4}}D(-\alpha)\right). \qquad (7)$$

For stabilization in the $q$ quadrature, the displacement is imaginary $\alpha = i\beta$ ($\beta \in \Re$), and for stabilization in the $p$ quadrature, the displacement is real $\alpha = \beta$. The four different Kraus operators associated with $\alpha = i\beta$ followed by $\alpha = \beta$, $M_j(\beta) =$ $M_{\pm}(\beta)M_{\pm}(i\beta)$ stabilize both quadratures—details in the supplementary materials (SM1).

We find that the $CD_4$ can create the same Kraus operators. The ancilla qudit is prepared in the initial state ($\bar{X}_4$ eigenstate) $|m=0\rangle$. The displacement is of $e^{i\pi/4}\sqrt{2}\beta$, and defining the measurement basis of the qudit ancilla by (detailed in SM2):

$$|j\rangle = \frac{1}{2}\sum_{s=0}^{3} e^{i\left(\beta^2 \cos(\pi s)+(\pi/2)\cos\left((s+j)\pi/2\right)\right)}|s\rangle. \tag{8}$$

The qudit measurement reveals two bits of information that can be assigned to the two quadratures. The Kraus operator applied to the oscillator is $M_j(\beta) = \langle j|CD_4\left(e^{i\pi/4}\sqrt{2}\beta\right)|m=0\rangle$. This Kraus operator can also be written as:

$$M_j(\beta) = \frac{1}{4}\sum_{s=0}^{3} e^{-\frac{i\pi}{2}\cos\left(\frac{\pi}{2}(s+j)\right)} e^{-(-1)^s i\beta^2} D\left(\sqrt{2}e^{i\frac{\pi}{4}(2s+1)}\beta\right). \tag{9}$$

For the case of a stabilizer measurement, $\beta = l/2 = \sqrt{\pi/2}$:

$$M_j\left(\sqrt{\pi/2}\right) = \frac{1}{4}\sum_{s=0}^{3} e^{-\frac{i\pi}{2}\left(\cos\left((s+j)\pi/2\right)+\cos(\pi s)\right)} D\left(e^{is\cdot\frac{2\pi}{4}}e^{\frac{i\pi}{4}}\sqrt{\pi}\right). \tag{10}$$

One possible Hermitian matrix that can be used for measuring the $|j\rangle$ basis for this case is:

$$\frac{1}{2}\begin{pmatrix} 3 & 2i & 1 & 0 \\ -2i & 3 & 0 & -1 \\ 1 & 0 & 3 & -2i \\ 0 & -1 & 2i & 3 \end{pmatrix}. \tag{11}$$

<u>Finite-energy GKP states</u>

Since the ideal measurement increases the energy of the GKP state, a correction process is needed to keep the envelope in the center and reduce the energy. For the finite-energy GKP case, we focus on the stabilization protocol of sharpen-trim (ST) used in [6, 18], shown in Figures 2a,c. Our results can also apply to the big-small-big (BSB) and small-big-small (SBS) protocols in [18], and we focus on the ST for simplicity. The ST protocol consists of two steps per quadrature: peak-sharpening to keep the oscillator state probability distribution peaked in $\hat{q} = 0 \bmod \sqrt{\pi}$ and $\hat{p} = 0 \bmod \sqrt{\pi}$, and envelope-trimming to prevent the overall envelope from drifting or expanding more than necessary, which reduces the mean number of photons. The peak-sharpening step consists of stabilizer measurement followed by a shift to correct the grid to $\hat{q} = 0 \bmod \sqrt{\pi}$ and $\hat{p} = 0 \bmod \sqrt{\pi}$. For our qudit protocol, the Kraus operator in this sharpen step is $M_j\left(\sqrt{\pi/2}\right)$ and the correction needed for the result $j$ is $D\left(\frac{\epsilon}{\sqrt{2}} e^{-\frac{i\pi}{4}(2j+1)}\right)$, where $\epsilon$ is defined as in [18], and it determines the width of the peaks. The envelope-trimming round consists of $CD_4$ with displacement of size $\epsilon$, followed by a shift of $D\left(\frac{l}{\sqrt{2}} e^{-\frac{i\pi}{4}(2\pi j+1)}\right)$ to recenter the envelope back to the center of phase space, restoring the symmetry of the grid.

The Kraus operators for the trim round are $M_{\pm}(\alpha)$ as in Eq. (7), with $\alpha = i\epsilon/2$ for the $q$ round or $\alpha = \epsilon/2$ for the $p$ round. For the qudit case, each round is applied on both quadratures simultaneously. Therefore, in the qudit case, the sharpen $p$ step is applied before the trim $q$ step. We show that the trim round almost commutes with the following sharpen round for the qubit protocol, allowing for switching between the sharpen $p$ step and the trim $q$ step for the qudit protocol. Sharpen-step in $p$ followed by trim-step in $q$ gives the following Kraus operator for the qubit case:

$$\frac{1}{4}\left(e^{(\pm1\pm1)i\frac{\pi}{4}} D(l) + e^{(\mp1\pm1)i\frac{\pi}{4}} I + e^{(\pm1\mp1)i\frac{\pi}{4}} e^{il\epsilon} D(-i\epsilon)D(l) + e^{(\mp1\mp1)i\frac{\pi}{4}} e^{il\epsilon} D(-i\epsilon)\right). \quad (12)$$

Trim-step in $q$ followed by sharpen-step in $p$ gives the following Kraus operator for the qubit case:

$$\frac{1}{4}\left(e^{(\pm1\pm1)i\frac{\pi}{4}} e^{-il\epsilon} D(l) + e^{(\mp1\pm1)i\frac{\pi}{4}} I + e^{(\pm1\mp1)i\frac{\pi}{4}} e^{il\epsilon} D(-i\epsilon)D(l) + e^{(\mp1\mp1)i\frac{\pi}{4}} D(-i\epsilon)\right). \quad (13)$$

In the second operation, the phase $e^{-il\epsilon}$ is added to the $D(l)$ part, and the phase $e^{il\epsilon}$ is removed in the $D(-i\epsilon)$ part. Since $e^{\pm il\epsilon}$ is a relatively small phase, it is possible to apply trim for both quadratures with the Kraus operator $M_j(\epsilon/2)$, as shown in Figure 2b. We choose $\epsilon = 0.1/\sqrt{2}$ for all the simulations. The lower $\epsilon$, the closer the two GKP manifolds of the two protocols are to each other.

The stabilization process of our proposed protocol is compared against the original qubit protocol [18]. A full stabilization round is completed after 2 CD operations for our qudit protocol and 4 CD operations for the original qubit protocol.

The expectation values approach 1 for the ideal GKP and remain lower for the finite-energy GKP. We compare the expectation values to the ideal stabilizers rather than the finite stabilizers, which provides a more robust comparison (since the finite version of the GKP manifold stabilized by our qudit protocol is slightly different from the qubit protocol). Figure 2d compares the expectation values for the stabilizers $\langle S_x\rangle, \langle S_y\rangle$ during both stabilization processes of a noiseless oscillator initialized with the vacuum state. Dashed (full) lines refer to the qubit (qudit) protocol. The qudit protocol shows improved stabilization over the qubit protocol, reaching expectation values closer to the ideal GKP faster.

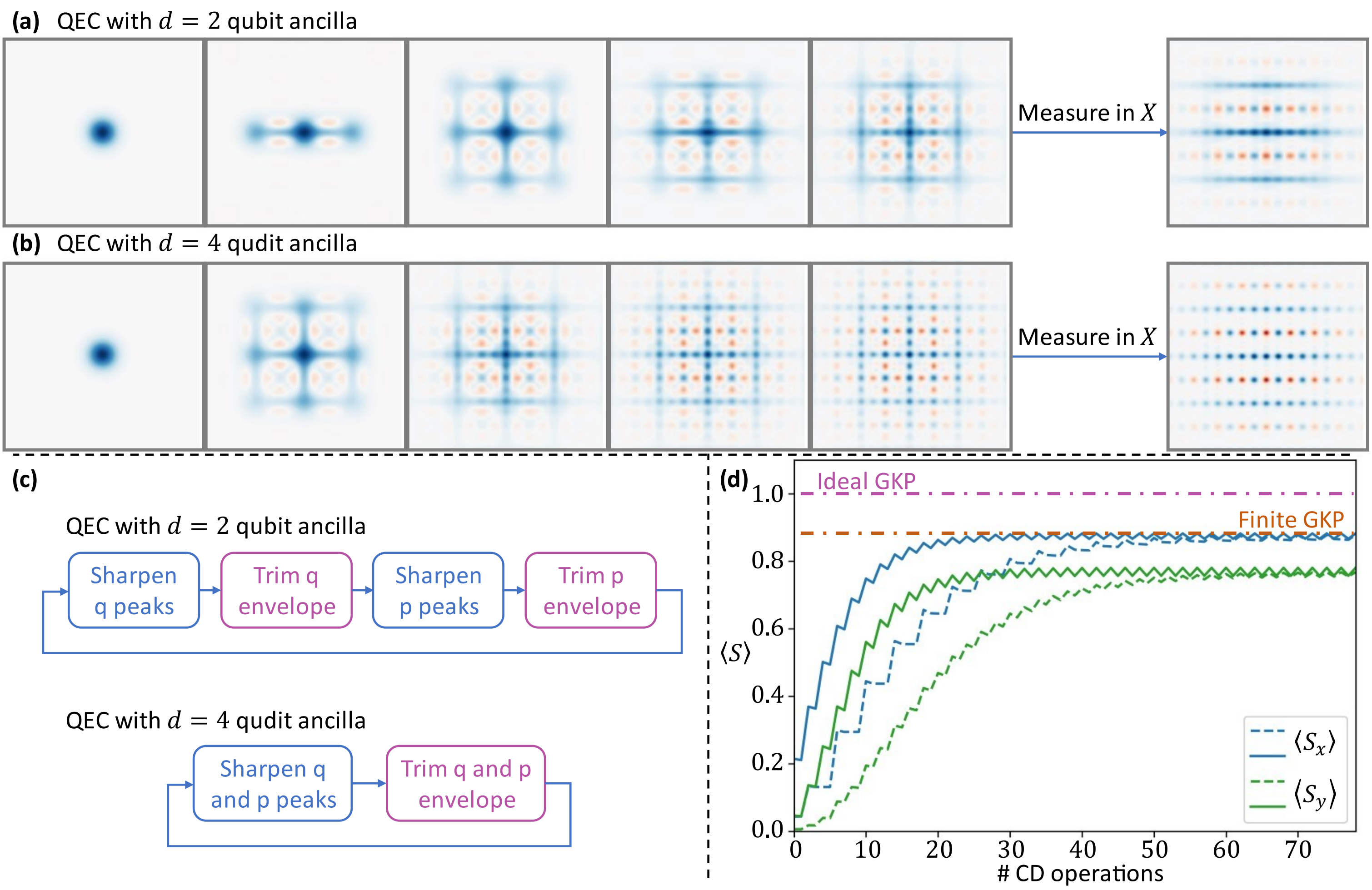

**Figure 2. Quantum error correction (QEC) and generation of GKP states using the generalized CD. (a)** Steps of the QEC protocol with qubit ancilla as presented in [6, 18]. The harmonic oscillator starts in a vacuum $|0\rangle$ state and converges to the GKP manifold. **(b)** Comparison to a protocol using a qudit ancilla with $d = 4$. The convergence to the GKP manifold is faster. **(c)** The protocols: each QEC round consists of sharpening steps that use CD with the grid's lattice constants as the displacements for squeezing the grid points and trimming steps that use CD with a small displacement

for limiting the Gaussian envelope of the grid. Using the generalized CD with $d = 4$, combines the two sharpen (trim) steps to a single CD sharpen (trim) operation. **(d)** Expectation values for the stabilizers as a function of the number of CD operations in a noiseless oscillator initialized with vacuum. Dashed (full) lines refer to the qubit (qudit) protocol.

**The resilience to oscillator errors improves for generalized CD**

Operating on both quadratures simultaneously and achieving symmetric stabilization reduces the inherent asymmetry in the previously proposed protocol. The asymmetry in that protocol causes asymmetrical accumulated errors in the GKP state due to stages with inherent asymmetry, as shown in Figure 2a. Depending on the implementation, reducing the number of steps may also reduce the time it takes to execute the protocols, which could assist in reducing the system's decoherence time and the error probability in a single QEC round, thus decreasing the probability of a logical error. Altogether, the generalized CD can help achieve faster QEC and reach break-even.

Figure 3 compares our proposed protocol against the qubit protocol to study the robustness of the logical information against typical errors. For this comparison, we used an error analysis similar to the one in [18], where the initial state is the approximated GKP state to which the stabilization protocol converges. We simulated both errors in the oscillator and in the ancilla. To simulate noise in the oscillator, we add idle times $\delta t$ (during which the oscillator evolves under a noise channel) between applications of the stabilization operations. The expectation values for the stabilizers are computed for two types of noise in the oscillator: photon loss at a rate $\kappa$ (Fig 3a) and dephasing at a rate $\kappa_\varphi$ (Fig 3b). The dotted lines represent GKP states without error correction. The stabilization protocols extend the lifespan of logical information, with the qudit protocol showing improvement compared to the qubit protocol.

To simulate noise in the ancilla, we apply the noise channel before applying the CD operations. The expectation values for the stabilizers are computed for three types of ancilla noise channels: $\bar{Z}_d$ noise $\mathcal{N}_Z(\rho_d) = p\bar{Z}_d\rho_d\bar{Z}_d^\dagger + (1-p)\rho_d$ (Fig 3c), $\bar{X}_d$ noise $\mathcal{N}_X(\rho_d) = p\bar{X}_d\rho_d\bar{X}_d^\dagger + (1-p)\rho_d$ (Fig 3d), and the depolarizing channel $\mathcal{N}_D(\rho_d) = p\mathbb{I}_d/d + (1-p)\rho_d$ (Fig 3e), where $\rho_d$ is the ancilla's state and $p$ is the error probability. The results show an advantage for the qudit protocol when the typical noise is $\mathcal{N}_X$ or when it is $\mathcal{N}_Z$ with low probability. The qubit protocol is superior when the typical noise is $\mathcal{N}_Z$ with high probability, or in the case of the depolarizing channel $\mathcal{N}_D$, in which the entire quantum information of the ancilla can be lost with probability $p$.

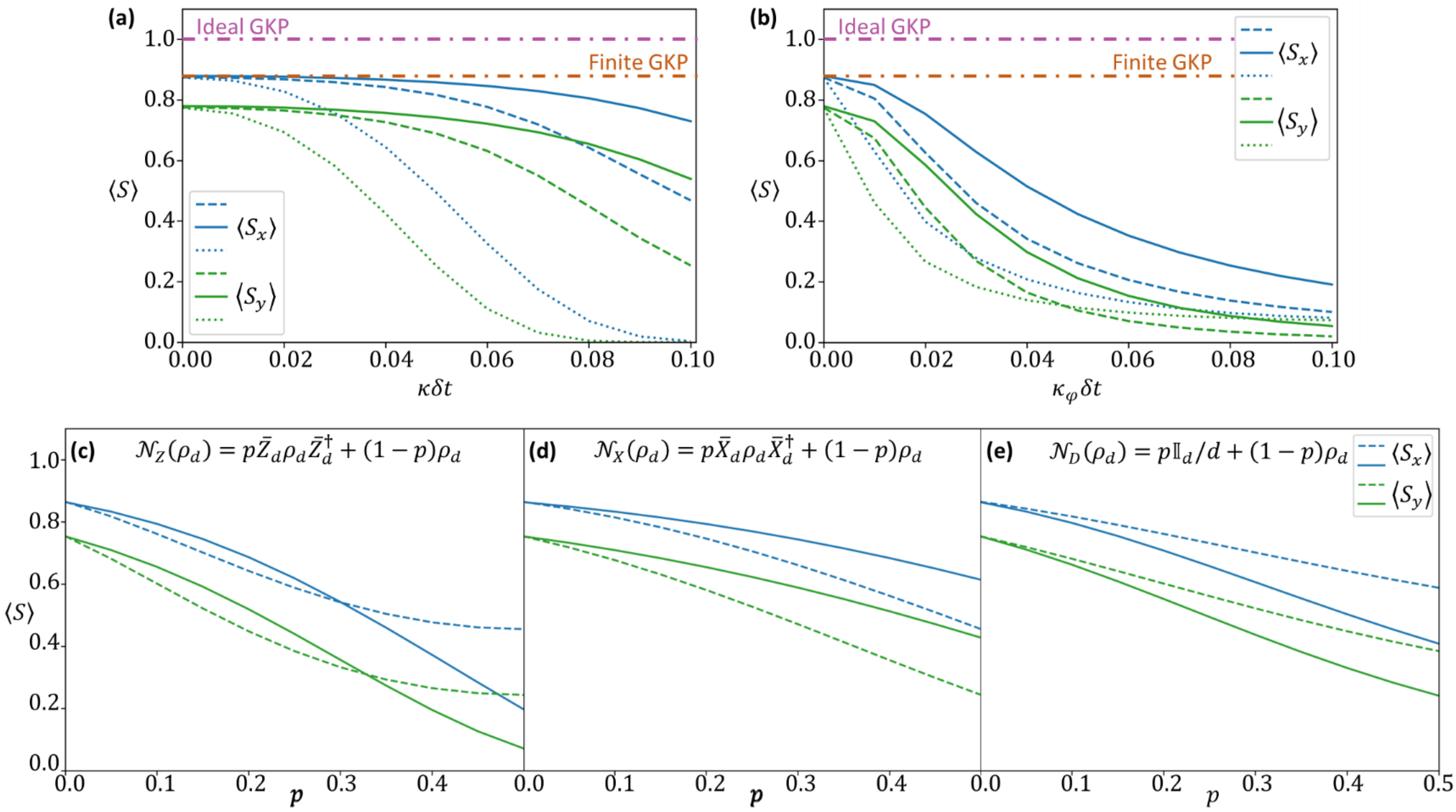

**Figure 3. Expectation values of the stabilizers as a function of noise.** The oscillator is initialized with the converged GKP state. Dashed (full) lines refer to the qubit (qudit) protocol, and the dotted lines refer to freely evolving GKP states with no QEC. **(a)** Photon loss in the oscillator at a rate $\kappa$. **(b)** Cavity dephasing in the oscillator at a rate $\kappa_\varphi$. **(c)** $\bar{Z}_d$ noise in the ancilla with probability $p$. **(d)** $\bar{X}_d$ noise in the ancilla with probability $p$. **(e)** Dephasing-channel noise in the ancilla with probability $p$.

**Hamiltonian implementation of generalized CDs**

This section proposes possible implementations of the qudit ancillas and their interactions with a harmonic oscillator, showing how to create the generalized CD operation. We first recall the case of a CD with a qubit ancilla. The most fundamental model for the interaction of a qubit with a harmonic oscillator is the quantum Rabi model [20], having $H_{\text{int}} = \hbar g X(a + a^\dagger)$, with $X$ the Pauli matrix of the qubit and $\hat{a}, \hat{a}^\dagger$ the annihilation and creation operators of the harmonic oscillator. Although the dynamics under this Hamiltonian do not directly create the CD operation, there exist specific regimes of deep-strong coupling [21, 22] for which the scattering matrix in the interaction picture is $e^{X(\alpha\hat{a}^\dagger-\alpha^*\hat{a})}$.

The qudit Rabi model [11] extends the qubit Rabi model to a qudit ancilla. Under the same approximation of deep-strong coupling, we get the scattering matrix $e^{\alpha\bar{X}_d{}^\dagger\hat{a}^\dagger-\alpha^*\bar{X}_d\hat{a}}$. Then, this scattering matrix naturally implements the generalized CD $D\left(\alpha\bar{X}_d{}^\dagger\right)$, conditioned on the $\hat{m}$ basis. Transforming to the $\hat{s}$ basis with the Fourier transform operator: $\mathcal{F}^\dagger\bar{X}_d{}^\dagger\mathcal{F}=\bar{Z}_d$ [11] results in a scattering matrix that equals the generalized conditional displacement operator $\mathrm{CD}_d(\alpha)$, as defined in Eq. (2), conditioned on the $\hat{s}$ basis.

Other approaches to implementing the generalized conditional displacement rely on encoding the qudit on a larger Hilbert space. In the following sub-sections, we examine such spaces and show implementations with different Rabi models and different scattering matrices without deep-strong coupling, each based on a different fundamental interaction.

Generalized CD operator using a planar rotor ancilla

The quantum rotor corresponds to the limit of a qudit with infinite levels [23], and it can be implemented using particles on a circle, the motion of electronic excitation in the periodic potential of a crystal, or a Cooper pair box [24]. The quantum rotor is defined by the conjugate operators of level number $\hat{N}$ and phase $\hat{\theta}$ [11]:

$$|N\rangle=\frac{1}{\sqrt{2\pi}}\int_{-\pi}^{\pi}d\theta e^{-i\theta N}|\theta\rangle,\;\;|\theta\rangle=\frac{1}{\sqrt{2\pi}}\sum_{N\in\mathbb{Z}}e^{i\theta N}|N\rangle. \tag{15}$$

The $|\theta\rangle$ states are not normalizable, as the quadrature states in harmonic oscillator $|q\rangle,|p\rangle$. The hopping operator in $|N\rangle$, $\bar{X}_{rot}|N\rangle=|N+1\rangle$, can be expressed as translation with the phase operator: $\bar{X}_{\mathrm{rot}}=e^{-i\hat{\theta}}$.

The rotor Rabi model [11] describes the simplest fundamental interaction between a rotor and a harmonic oscillator of transitions in $\hat{N}$ while also absorbing or emitting a photon. In this case, the scattering matrix for the interaction Hamiltonian is in the form of $e^{\alpha e^{i\hat{\theta}}\hat{a}^\dagger-\alpha^* e^{-i\hat{\theta}}\hat{a}}$. Since the quantum rotor is the limit of a qudit with $d\rightarrow$ $\infty$, this scattering matrix does not contain the counter-rotating terms. Therefore, unlike in the qudit case, this interaction can be achieved without deep-strong coupling. This scattering matrix implements the generalized conditional displacement operator conditioned on the phase: $D\left(\alpha e^{i\hat{\theta}}\right) \equiv \mathrm{CD_{rot}}(\alpha) = \int_{-\pi}^{\pi} d\theta |\theta\rangle\langle\theta| \otimes D\left(\alpha e^{i\theta}\right)$.

The $d$-legged cat state $|C_d^N(\alpha)\rangle$ as defined in Eq. (3), has a generalized parity of $N$ mod $d$, therefore in the limit of $d \to \infty$, it is the Fock state $|N\rangle$ (SM3). As shown in Eq. (6) with the $d$-legged cat state created by $\mathrm{CD}_d(\alpha)$, applying $\mathrm{CD_{rot}}(\alpha)$ on the initial state of the ancilla rotor in the state $|N\rangle$ and measuring $\hat{N}$ on the ancilla rotor will apply the following Kraus operator on the harmonic oscillator:

$$\langle N'|\mathrm{CD_{rot}}(\alpha)|N\rangle = \frac{1}{2\pi}\int_{-\pi}^{\pi} d\theta e^{i\theta(N'-N)} D\left(\alpha e^{i\theta}\right). \tag{16}$$

Therefore, when the initial state of the oscillator is the vacuum $|0\rangle$ state, the oscillator result state is $\left|C_{d\to\infty}^{N'-N}(\alpha)\right\rangle$, which is equal to the Fock state $|N' - N\rangle$ (details in SM3).

The planar rotor can encode a logical qudit in the rotor GKP encoding [24], providing redundancy that enables an error-robust qudit. The rotor GKP can encode a qudit by defining the logical states as $|s\rangle_L = |\theta = 2\pi s/d\rangle$, resulting in the hopping operator $e^{i\hat{\theta}}$ to be the logical $\bar{Z}_d$ operator since $e^{i\hat{\theta}}|s\rangle_L = e^{i2\pi s/d}|s\rangle_L = \bar{Z}_d|s\rangle_L$. This choice of encoding simulates the $\mathrm{CD}_d(\alpha)$ operator with the scattering matrix:

$$D\left(\alpha e^{i\hat{\theta}}\right) = D(\alpha \bar{Z}_d) = \mathrm{CD}_d(\alpha). \tag{17}$$

Since the logical states are not normalizable, we can define the finite version of the rotor-GKP code with a Gaussian envelope with variance $\sigma^2$ and centered around $n$ (Figure 4a):

$$|s\rangle_L^{n,\sigma^2} = \frac{1}{\sqrt{\sigma\sqrt{2\pi}}}\sum_{N\in\mathbb{Z}} e^{-\frac{(N-n)^2}{4\sigma^2}} e^{\frac{i2\pi s N}{d}} |N\rangle. \tag{18}$$

<u>Generalized CD operator using a beam splitter with a cat code ancilla</u>

The next scattering matrix we explore is the beam splitter: $e^{\theta \hat{a}'\hat{a}^\dagger - \theta^* \hat{a}'^\dagger \hat{a}}$. In this case, the ancilla system is also a harmonic oscillator, with the ladder operators $\hat{a}', \hat{a}'^\dagger$. There are many ways to encode a qudit on a harmonic oscillator in bosonic codes, and we discussed here the two prominent codes – the cat code and the GKP code. We show that the beam-splitter interaction approximates the generalized CD when the ancilla encodes a qudit in the cat code [13]. In this encoding, the logical states are the coherent states $|s\rangle_L = |\alpha e^{i2\pi s/d}\rangle$ and the conjugate basis cat states as defined in Eq. (3), $|m\rangle_L = |C_d^m\rangle$. The operator $\frac{1}{\alpha}\hat{a}'$ can approximate the logical $\bar{Z}_d$ operator, since $\frac{1}{\alpha}\hat{a}'|s\rangle_L = e^{i2\pi s/d}|\alpha e^{i2\pi s/d}\rangle = \bar{Z}_d|s\rangle_L$. The operator $\frac{1}{\alpha^*}\hat{a}'^\dagger$ is not exactly the logical $\bar{Z}_d^\dagger$ operator, but approximates it better as $\alpha$ increases, as shown in figure 4b. This choice of encoding simulates the $\mathrm{CD}_d$ operator with the beam splitter operator:

$$e^{\theta\hat{a}'\hat{a}^\dagger-\theta^*\hat{a}'^\dagger\hat{a}} \approx e^{\theta\alpha\bar{Z}_d\hat{a}^\dagger-\theta^*\alpha^*\bar{Z}_d{}^\dagger\hat{a}} = \mathrm{CD}_d(\theta\alpha). \tag{19}$$

<u>Generalized CD operator using a permutation-invariant code with a spin ancilla</u>

The final interaction we explore is the Dicke model [25], a fundamental model of the interaction between a harmonic oscillator and a set of indistinguishable two-level systems, or equivalently, a single spin- $N/2$ system. Under the rotating-wave approximation, the counter-rotating terms are omitted and the Dicke model reduces to the Tavis–Cummings model [26]. In the Tavis–Cummings model, the scattering matrix for the interaction Hamiltonian is $e^{\alpha S_-\hat{a}^\dagger-\alpha^* S_+\hat{a}}$, where $S_\pm$are the spin ladder operators.

We show that the Tavis–Cummings interaction approximates the generalized CD when we encode the qudit on the spin ancilla using the permutation-invariant codes [27]. This encoding of the qudit in the Dicke symmetric states is invariant to permutations on the two-level systems. We find the specific permutation-invariant code that corresponds to the Tavis–Cummings interaction, where the logical states are the spin-coherent states $|s\rangle_L = \left|\theta = \frac{\pi}{2}, \varphi = \frac{2\pi s}{d}\right\rangle$. The operator $\frac{2}{N}S_-$ can approximate the logical $\bar{Z}_d$ operator (see details in SM 3):

$$\frac{2}{N}S_-|s\rangle_L \approx e^{\frac{i2\pi s}{d}}|s\rangle_L = \bar{Z}_d|s\rangle_L. \tag{21}$$

The approximation improves as $N \to \infty$, as shown in figure 4c. This choice of encoding simulates the $\mathrm{CD}_d$ operator with the spin scattering matrix:

$$e^{\alpha S_-\hat{a}^\dagger-\alpha^* S_+\hat{a}} \approx e^{\frac{N}{2}\left(\alpha\bar{Z}_d\hat{a}^\dagger-\alpha^*\bar{Z}_d^\dagger\hat{a}\right)} = \mathrm{CD}_d(\alpha N/2). \tag{22}$$

Ref. [12] suggested an implementation of GKP generation using an atomic ensemble ancilla, which implements a different generalization of the CD operation— extending from qubit to a higher spin, replacing the Heisenberg-Weyl qudit operators we used here with the spin operators.

This approximation of the generalized CD in a Dicke system could have important consequences, for example laying a path for the creation of approximated GKP states in the Dicke symmetric-state Hilbert space. Such approximated GKP states can then be transferred to desired traveling photonic states via spontaneous emission, once selecting the coupling parameters to ensure single-mode coupling [28, 29]. The prospects of this approach for creating traveling GKP states could be especially intriguing in the optical domain, where such states are hard to create with sufficient fidelity, but new recent platforms now enable collective emission from Dicke states [30–33].

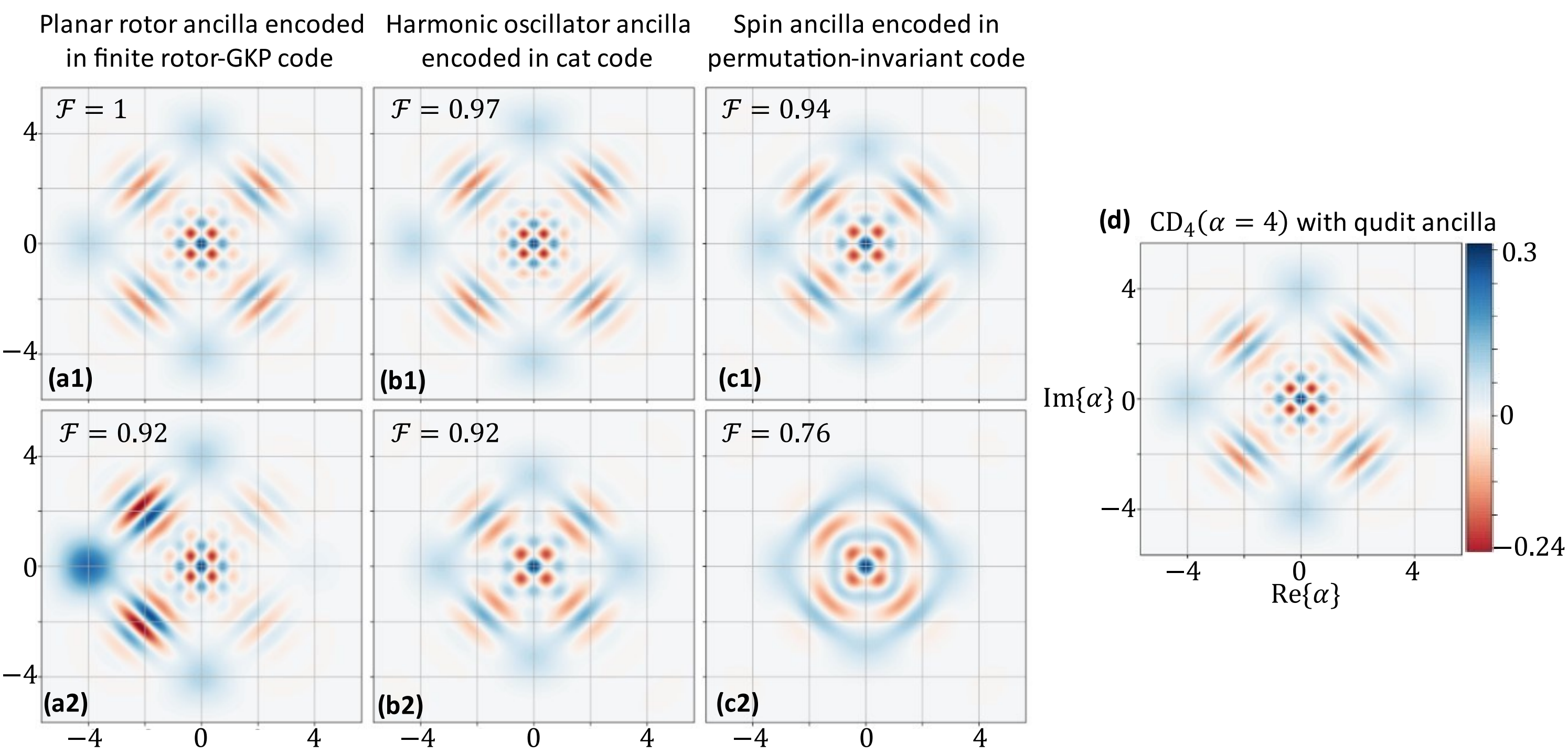

**Figure 4. Comparison of implementations of the generalized CD operator.** We estimate the quality of each implementation of the generalized CD by the fidelity of the resulting 4-legged cat state. **(a)** Planar rotor ancilla encoded in finite rotor-GKP code with variance $\sigma^2 = 4$ centered around $n = 20$ (a1) and with variance $\sigma^2 = 1$ centered around $n = 30$ (a2). **(b)** Harmonic oscillator ancilla encoded in cat code with $\alpha = 4$ (b1) and $\alpha = 3$ (b2). **(c)** Spin-$N/2$ ancilla encoded in a permutation-invariant code with $N = 50$ (c1) and $N = 19$ (c2). **(d)** Resulting 4-legged cat state from the ideal generalized CD implementation with qudit ancilla.

The three encodings we used for the planar rotor, harmonic oscillator, and Dicke spin can all be considered as finite approximations of the non-normalizable Pegg-Barnett phase states [34]. For the planar rotor, the approximate Pegg-Barnett states are the normalized $|\theta\rangle$ states. For the harmonic oscillator, they are the bosonic rotation codes, such as the binomial or cat code [12] For the spin-$N/2$, what approximates the Pegg-Barnett states are the permutation-invariant codes [27] that take the same binomial code [1] form in terms of excitation numbers.

## Experimental perspective and outlook

We conclude by discussing possible physical platforms for implementing the generalized CD. One possibility is a transmon qudit coupled to a microwave cavity in a circuit QED. The cavity mode is dispersively coupled to the transmon qudit ancilla, which induces anharmonicity depending on its state. The induced anharmonic frequency shifts are small, in the kHz-MHz range. In such a system, a generalized CD can be applied as an extension of more conventional implementations with two-level ancilla qubits. Specifically, shaping a coherent-control pulse could create a complex excitation with a few nodes and peaks of required amplitudes and phases at a few selected frequencies [35]. These frequencies should correspond to the anharmonic shifts of multiple energy levels. It seems that this case would be limited in practice to about four levels [36], which is already sufficient to demonstrate the predictions of our work.

A second possible implementation is using quantum rotor systems, which are created naturally using free electrons interacting with laser fields [37–40]. As recent experiments in electron microscopes show, the electrons achieve especially strong interactions by propagating parallel to (grazing) elongated photonic waveguides or cavities [41–45]. Such interactions could implement the generalized CD in optical frequencies, as first proposed for a particular case in [15, 16].

It is now rare to discover fundamentally new physical platforms for implementations of entangling gates such as CNOT in discrete-variable systems. However, in continuous-variable systems, discoveries of new possible platforms are still being made for entangling gates such as CD [15, 16, 28, 29]. Our work shows that implementations of generalized CD can arise from one of several fundamental interactions, which can be classified by four types of Hilbert spaces, each describing a different ancilla coupled to the harmonic oscillator. Each of these interactions can form this basic building block for universal computation and QEC with GKP states. We also showed that the generalized CD operator of any size can be used to create the $d$-legged cat states (Figure 1c), which are rotation symmetric bosonic code states [14]. This code is quantum error correctable for Fock space shifts and phase space rotations. Such discoveries motivate a wider search of new physical interactions for continuous-variable quantum information processing.

## Acknowledgments

This research project was supported under the flagship research project QUBIT by the Helen Diller Quantum Center at the Technion. The work was also supported by the Israel Science Foundation, Grant No. 830/19. S.E.-H. acknowledges financial support from the Ariane de Rothschild Women Doctoral Program. R.R. acknowledges financial support from the ADAMS Fellowship Program of the Israel Academy of Sciences and Humanities.

## References

[1] M. H. Michael et al., "New Class of Quantum Error-Correcting Codes for a Bosonic Mode," Phys. Rev. X, vol. 6, no. 3, p. 031006, Jul. 2016, https://doi.org/10.1103/PhysRevX.6.031006.

[2] P. T. Cochrane, G. J. Milburn, and W. J. Munro, "Macroscopically distinct quantum-superposition states as a bosonic code for amplitude damping," Phys. Rev. A, vol. 59, no. 4, pp. 2631–2634, Apr. 1999, https://doi.org/10.1103/PhysRevA.59.2631.

[3] D. Gottesman, A. Kitaev, and J. Preskill, "Encoding a qubit in an oscillator," Phys. Rev. A, vol. 64, no. 1, p. 012310, Jun. 2001, https://doi.org/10.1103/PhysRevA.64.012310.

[4] Y. Liu et al., "Hybrid Oscillator-Qubit Quantum Processors: Instruction Set Architectures, Abstract Machine Models, and Applications," Aug. 05, 2024, arXiv: arXiv:2407.10381. https://doi.org/10.48550/arXiv.2407.10381.

[5] C. Flühmann, T. L. Nguyen, M. Marinelli, V. Negnevitsky, K. Mehta, and J. P. Home, "Encoding a qubit in a trapped-ion mechanical oscillator," Nature, vol. 566, no. 7745, Art. no. 7745, Feb. 2019, https://doi.org/10.1038/s41586-019-0960-6.

[6] P. Campagne-Ibarcq et al., "Quantum error correction of a qubit encoded in grid states of an oscillator," Nature, vol. 584, no. 7821, Art. no. 7821, Aug. 2020, https://doi.org/10.1038/s41586-020-2603-3.

[7] V. V. Sivak et al., "Real-time quantum error correction beyond break-even," Nature, vol. 616, no. 7955, Art. no. 7955, Apr. 2023, https://doi.org/10.1038/s41586-023-05782-6.

[8] A. Eickbusch et al., "Fast universal control of an oscillator with weak dispersive coupling to a qubit," Nat. Phys., vol. 18, no. 12, Art. no. 12, Dec. 2022, https://doi.org/10.1038/s41567-022-01776-9.

[9] B. M. Terhal, J. Conrad, and C. Vuillot, “Towards scalable bosonic quantum error correction,” Quantum Sci. Technol., vol. 5, no. 4, p. 043001, Jul. 2020, https://doi.org/10.1088/2058-9565/ab98a5.
[10] A. Asadian, P. Erker, M. Huber, and C. Klöckl, “Heisenberg-Weyl Observables: Bloch vectors in phase space,” Phys. Rev. A, vol. 94, no. 1, p. 010301, Jul. 2016, https://doi.org/10.1103/PhysRevA.94.010301.
[11] V. V. Albert, S. Pascazio, and M. H. Devoret, “General phase spaces: from discrete variables to rotor and continuum limits,” J. Phys. A: Math. Theor., vol. 50, no. 50, p. 504002, Nov. 2017, https://doi.org/10.1088/1751-8121/aa9314.
[12] K. R. Motes, B. Q. Baragiola, A. Gilchrist, and N. C. Menicucci, “Encoding qubits into oscillators with atomic ensembles and squeezed light,” Phys. Rev. A, vol. 95, no. 5, p. 053819, May 2017, https://doi.org/10.1103/PhysRevA.95.053819.
[13] L. Li, C.-L. Zou, V. V. Albert, S. Muralidharan, S. M. Girvin, and L. Jiang, “Cat Codes with Optimal Decoherence Suppression for a Lossy Bosonic Channel,” Phys. Rev. Lett., vol. 119, no. 3, p. 030502, Jul. 2017, https://doi.org/10.1103/PhysRevLett.119.030502.
[14] A. L. Grimsmo, J. Combes, and B. Q. Baragiola, “Quantum Computing with Rotation-Symmetric Bosonic Codes,” Phys. Rev. X, vol. 10, no. 1, p. 011058, Mar. 2020, https://doi.org/10.1103/PhysRevX.10.011058.
[15] R. Dahan, G. Baranes, A. Gorlach, R. Ruimy, N. Rivera, and I. Kaminer, “Creation of Optical Cat and GKP States Using Shaped Free Electrons,” Phys. Rev. X, vol. 13, no. 3, p. 031001, Jul. 2023, https://doi.org/10.1103/PhysRevX.13.031001.
[16] G. Baranes et al., “Free-electron interactions with photonic GKP states: Universal control and quantum error correction,” Phys. Rev. Research, vol. 5, no. 4, p. 043271, Dec. 2023, https://doi.org/10.1103/PhysRevResearch.5.043271.
[17] J. Zak, “Finite Translations in Solid-State Physics,” Phys. Rev. Lett., vol. 19, no. 24, pp. 1385–1387, Dec. 1967, https://doi.org/10.1103/PhysRevLett.19.1385.
[18] B. Royer, S. Singh, and S. M. Girvin, “Stabilization of Finite-Energy Gottesman-Kitaev-Preskill States,” Phys. Rev. Lett., vol. 125, no. 26, p. 260509, Dec. 2020, https://doi.org/10.1103/PhysRevLett.125.260509.
[19] B. M. Terhal and D. Weigand, “Encoding a qubit into a cavity mode in circuit QED using phase estimation,” Phys. Rev. A, vol. 93, no. 1, p. 012315, Jan. 2016, https://doi.org/10.1103/PhysRevA.93.012315.
[20] I. I. Rabi, “On the Process of Space Quantization,” Phys. Rev., vol. 49, no. 4, pp. 324–328, Feb. 1936, https://doi.org/10.1103/PhysRev.49.324.
[21] A. Frisk Kockum, A. Miranowicz, S. De Liberato, S. Savasta, and F. Nori, “Ultrastrong coupling between light and matter,” Nat Rev Phys, vol. 1, no. 1, Art. no. 1, Jan. 2019, https://doi.org/10.1038/s42254-018-0006-2.
[22] J. Hastrup, K. Park, J. B. Brask, R. Filip, and U. L. Andersen, “Measurement-free preparation of grid states,” npj Quantum Inf, vol. 7, no. 1, Art. no. 1, Jan. 2021, https://doi.org/10.1038/s41534-020-00353-3.
[23] B. W. Shore and J. H. Eberly, “Analytic approximations in multi-level excitation theory,” Optics Communications, vol. 24, no. 1, pp. 83–88, Jan. 1978, https://doi.org/10.1016/0030-4018(78)90272-9.
[24] V. V. Albert, J. P. Covey, and J. Preskill, “Robust Encoding of a Qubit in a Molecule,” Phys. Rev. X, vol. 10, no. 3, p. 031050, Sep. 2020, https://doi.org/10.1103/PhysRevX.10.031050.
[25] R. H. Dicke, “Coherence in Spontaneous Radiation Processes,” Phys. Rev., vol. 93, no. 1, pp. 99–110, Jan. 1954, https://doi.org/10.1103/PhysRev.93.99.
[26] M. Tavis and F. W. Cummings, “Exact Solution for an $N$-Molecule---Radiation-Field Hamiltonian,” Phys. Rev., vol. 170, no. 2, pp. 379–384, Jun. 1968, https://doi.org/10.1103/PhysRev.170.379.
[27] Y. Ouyang, “Permutation-invariant qudit codes from polynomials,” Linear Algebra and its Applications, vol. 532, pp. 43–59, Nov. 2017, https://doi.org/10.1016/j.laa.2017.06.031.

[28] O. Tziperman et al., “Spontaneous emission from correlated emitters,” Jun. 20, 2023, arXiv: arXiv:2306.11348. https://doi.org/10.48550/arXiv.2306.11348.
[29] N. Gutman, A. Gorlach, O. Tziperman, R. Ruimy, and I. Kaminer, “Universal Control of Symmetric States Using Spin Squeezing,” Dec. 03, 2023, arXiv: arXiv:2312.01506. https://doi.org/10.48550/arXiv.2312.01506.
[30] D. M. Lukin et al., “Two-Emitter Multimode Cavity Quantum Electrodynamics in Thin-Film Silicon Carbide Photonics,” Phys. Rev. X, vol. 13, no. 1, p. 011005, Jan. 2023, https://doi.org/10.1103/PhysRevX.13.011005.
[31] A. Tiranov et al., “Collective super- and subradiant dynamics between distant optical quantum emitters,” Science, vol. 379, no. 6630, pp. 389–393, Jan. 2023, https://doi.org/10.1126/science.ade9324.
[32] G. Ferioli, S. Pancaldi, A. Glicenstein, D. Clement, A. Browaeys, and I. Ferrier-Barbut, “Non-Gaussian correlations in the steady-state of driven-dissipative clouds of two-level atoms,” Feb. 13, 2024, arXiv: arXiv:2311.13503. https://doi.org/10.48550/arXiv.2311.13503.
[33] C. Liedl, F. Tebbenjohanns, C. Bach, S. Pucher, A. Rauschenbeutel, and P. Schneeweiss, “Observation of Superradiant Bursts in a Cascaded Quantum System,” Phys. Rev. X, vol. 14, no. 1, p. 011020, Feb. 2024, https://doi.org/10.1103/PhysRevX.14.011020.
[34] S. M. Barnett and D. T. Pegg, “Phase in quantum optics,” J. Phys. A: Math. Gen., vol. 19, no. 18, p. 3849, Dec. 1986, https://doi.org/10.1088/0305-4470/19/18/030.
[35] A. A. Diringer, E. Blumenthal, A. Grinberg, L. Jiang, and S. Hacohen-Gourgy, “Conditional not displacement: fast multi-oscillator control with a single qubit”, Phys. Rev. X 14, 011055, March 2024, https://doi.org/10.1103/PhysRevX.14.011055.
[36] M. J. Peterer et al., “Coherence and Decay of Higher Energy Levels of a Superconducting Transmon Qubit,” Phys. Rev. Lett., vol. 114, no. 1, p. 010501, Jan. 2015, https://doi.org/10.1103/PhysRevLett.114.010501.
[37] B. Barwick, D. J. Flannigan, and A. H. Zewail, “Photon-induced near-field electron microscopy,” Nature, vol. 462, no. 7275, Art. no. 7275, Dec. 2009, https://doi.org/10.1038/nature08662.
[38] S. T. Park, M. Lin, and A. H. Zewail, “Photon-induced near-field electron microscopy (PINEM): theoretical and experimental,” New J. Phys., vol. 12, no. 12, p. 123028, Dec. 2010, https://doi.org/10.1088/1367-2630/12/12/123028.
[39] F. J. García de Abajo, A. Asenjo-Garcia, and M. Kociak, “Multiphoton Absorption and Emission by Interaction of Swift Electrons with Evanescent Light Fields,” Nano Lett., vol. 10, no. 5, pp. 1859–1863, May 2010, https://doi.org/10.1021/nl100613s.
[40] O. Reinhardt, C. Mechel, M. Lynch, and I. Kaminer, “Free-Electron Qubits,” Annalen der Physik, vol. 533, no. 2, p. 2000254, 2021, https://doi.org/10.1002/andp.202000254.
[41] R. Dahan et al., “Resonant phase-matching between a light wave and a free-electron wavefunction,” Nat. Phys., vol. 16, no. 11, Art. no. 11, Nov. 2020, https://doi.org/10.1038/s41567-020-01042-w.
[42] R. Dahan et al., “Imprinting the quantum statistics of photons on free electrons,” Science, vol. 373, no. 6561, p. eabj7128, Aug. 2021, https://doi.org/10.1126/science.abj7128.
[43] J.-W. Henke et al., “Integrated photonics enables continuous-beam electron phase modulation,” Nature, vol. 600, no. 7890, Art. no. 7890, Dec. 2021, https://doi.org/10.1038/s41586-021-04197-5.
[44] A. Feist et al., “Cavity-mediated electron-photon pairs,” Science, vol. 377, no. 6607, pp. 777–780, Aug. 2022, https://doi.org/10.1126/science.abo5037.
[45] Y. Adiv et al., “Observation of 2D Cherenkov Radiation,” Phys. Rev. X, vol. 13, no. 1, p. 011002, Jan. 2023, https://doi.org/10.1103/PhysRevX.13.011002.